# Self-sensing with hollow cylindrical transducers for histotripsy enhanced aspiration mechanical thrombectomy applications

Li Gong[a,b,*], Alex R. Wright[b], Kullervo Hynynen[a,b,c], David E. Goertz[a,b]

a Department of Medical Biophysics, University of Toronto, Toronto, M5S 1A1, Canada
b Sunnybrook Research Institute, 2075 Bayview Avenue, M4N 3M5, Toronto, Canada
c Institute of Biomaterials and Biomedical Engineering, University of Toronto, Toronto, Canada

*Corresponding author at: Physical Sciences Platform, Sunnybrook Research Institute, Sunnybrook Health Sciences Centre, 2075 Bayview Avenue, Rm A665a, Toronto, ON, M4N 3M5, Canada. E-mail address: li.gong@mail.utoronto.ca (L. Gong).

## Abstract

To address existing challenges with intravascular mechanical thrombectomy devices, a novel ultrasound-enhanced aspiration approach is being developed to mechanically degrade clots using cavitation. This method employs standing waves within a mm-scale hollow cylindrical transducer to generate high pressures sufficient to perform histotripsy on clots situated within the transducer lumen and generate substantial lesions. The objective of this study is to assess the feasibility of self-sensing cavitation detection by analyzing voltage signals across the transducer during treatment pulses. External ultrasound imaging of the transducer lumen validated cavitation detection. Impedance was also altered by the presence of clot material within the lumen. Experiments varying the driving voltage in water-filled lumens demonstrated changes in the relative amplitudes of the envelopes of the pulse body and ringdown portions of the voltage signals above the cavitation threshold, as well as changes in the spectral domain. In particular both broadband and ultraharmonic signals showed an increase in amplitude above the cavitation threshold. Similar temporal and spectral voltage signal changes in the presence of cavitation were also observed when treating clots within the lumen. This work demonstrates a highly sensitive method for detecting cavitation within the lumen, enabling monitoring with readily acquired signals without additional sensors in the catheter configuration. These findings hold significant potential for improving the efficacy of ultrasound-enhanced aspiration thrombectomy procedures.

# Introduction

Thrombotic occlusions significantly contribute to mortality and morbidity globally, particularly in cases of myocardial infarction [1–3], ischemic stroke [4], pulmonary embolism (PE) [5], and deep venous thrombosis (DVT) [6]. Given the widespread occurrence and substantial health impact of these large vessel thrombotic occlusions, various treatment methods have been developed. Traditionally, thrombolytic enzymes have been a primary treatment option, working by breaking down the fibrin network and thus the structural integrity of clots. More recently, catheter-based mechanical thrombectomy techniques have become increasingly common in standard clinical practice to remove acute thrombotic occlusions in large vessels. Mechanical thrombectomy procedures are employed in the treatment of stroke [7–11], pulmonary embolism [5,12–14] and peripheral vessels [6,15,16]. One of the most widely used mechanical thrombectomy methods is aspiration, which operates by applying suction to extract clot material through the lumen of a hollow catheter. While this approach has significantly impacted the treatment of patients, in many circumstances it is unsuccessful or performs sub-optimally [17,18] due to thrombus clogging within the catheter or 'corking' at the catheter tip [19,20]. Therefore, there is a recognized need to improve the performance of aspiration mechanical thrombectomy devices, particularly their ability to ingest more challenging clots (e.g., stiffer or larger volume clots).

We are developing a novel ultrasound-enhanced aspiration approach, which involves the placement of a hollow cylindrical transducer (HCT) at the catheter tip. The premise is to perform histotripsy on clot material within the HCT lumen, degrading the clot's mechanical integrity and thereby facilitating its ingestion further into the catheter. Histotripsy is an established ultrasound method that employs cavitation clouds to fractionate targeted regions of tissue into an acellular liquid [21–24]. The approach involves using very high-pressure ultrasound to generate violent bubble clouds within the transducer focal region. There are several general classes of histotripsy, determined by the pulse lengths employed: extrinsic histotripsy typically uses ~1-2 µs pulses and requires the highest pressure levels; shock scattering histotripsy uses 5-20 µs pulses; and 'boiling' histotripsy involves 1-100 ms pulses and requires the lowest pressure level due to pulse-scale thermal elevations. With few exceptions [25–27], large aperture extracorporeal spherically focused transducer configurations are used to achieve the pressure levels required to create cavitation clouds.

With the proposed aspiration implementation, sufficiently high pressures must be achieved within the lumen of HCTs that are of a scale compatible with intravascular catheters. In prior work [28], we demonstrated that cavitation clouds could be generated within a radially polarized HCT (3.3/2.5 mm outer/inner diameter) when operating within its 'thickness' mode resonant frequency bandwidth. The associated transducer vibrations produce cylindrical waves and a high-pressure zone along the transducer axis. In particular, when operating at a standing wave frequency, constructive interference can increase the effective focal gain, resulting in increased internal pressures (estimated > 20 MPa). In this work, cavitation within the lumen was confirmed with a hydrophone situated external to the transducer lumen, and high-frequency (40 MHz) ultrasound imaging. The pulse lengths employed (10 and 100 μs) were sufficiently long to permit the development of standing wave. This was supported by hydrophone data that revealed pressure build up at the outset of electrical pulse stimulation that was associated with the development of the interference pattern, and a decay (ringdown) following the end of the pulse associated with the dissipation of internal reflections. In more recent work [29], we demonstrated the feasibility of performing histotripsy within clots situated in the HCT lumen.

An important element of any cavitation-mediated therapeutic ultrasound approach is monitoring and controlling the level of cavitation present. Cavitation monitoring and control methods have been widely investigated in the setting of microbubble-mediated drug delivery [30–33]. Generally, these involve using receive transducers, separate from the transmit transducers generating the therapy beam, to passively detect and, in some cases, spatially map cavitation emissions arising from acoustic stimulation by the therapy pulses. In histotripsy, the primary approach to cavitation monitoring to date has involved situating an ultrasound imaging transducer within the therapy transducer aperture. B-Scan ultrasound imaging is then used to 'actively' (pulse-echo) detect bubble clouds that form during the treatment process. The imaging information is used for targeting purposes, confirming the formation and extent of lesions, and potentially adjusting the transmit conditions to ensure sufficient cavitation activity. More recently, efforts are underway to develop transmit-receiver arrays compatible with short pulse (extrinsic) histotripsy. In [34,35], a subset of elements was configured with both transmit and receive capabilities, enabled by the short duration of the transmitted pulses relative to the propagation time to regions of interest, resulting in time separation between the transmit and receive signals.

For the proposed ultrasound-enhanced aspiration approach, as with conventional histotripsy, the ability to detect and monitor induced cavitation will be necessary to ensure that treatments are occurring and to potentially adjust transmit parameters as necessary. As the intended application areas are in the setting of stroke, PE, and the peripheral vasculature, using a separate B-scan imaging transducer to monitor cavitation within the HCT lumen is not viable. Further, due to the pulse lengths employed relative to the lumen size, there will not be temporal separation between the transmit and receive signals, as is the case in [28]. In the present study, we will investigate the use of voltage signals measured across the transmit transducer to monitor cavitation induced within the lumen of the HCT. In the setting of boiling histotripsy conditions, using a single-element large aperture spherically focused transducer, there have been several reports of fluctuations in the voltage or power across the transducer after the onset of cavitation within the focal region [36–42]. These fluctuations result from the signals across the transducer being a superposition of the applied transmit voltage and the arrival of emissions and reflections/scattering from the focal cavitation cloud. Outside the biomedical ultrasound field, particularly in the area of sonochemistry, there have been numerous reports of examining of measured transducer signals (voltage and current) to monitor cavitation with long duration pulses (>100 ms). This work has been carried out for acoustic horn or larger-scale cylindrical focusing configurations. It has been shown that the measured signals can detect the presence of inertial cavitation, harmonics and subharmonics [43,44]. This is often referred to as 'self-sensing' due to the dual use of the transducer as a transmitter and receiver, with the target cavitation signals being concurrent with the transmission stimulation signal.

Here we investigate the use of self-sensing approaches to detect and monitor cavitation within an intravascular scale HCT. Following the conditions employed in recent work (10 and 100 μs pulses) [28], signals measured across the HCT will be assessed over a range of applied voltages from sub-cavitation to cavitation levels. The temporal (pulse-body and ringdown) and spectral aspects of the pulses will be assessed. The presence of cavitation will be confirmed with high frequency ultrasound imaging. The majority of experiments will be conducted with water-filled lumens, followed by a proof-of-principle experiment with thrombus-filled lumens. Additionally, electrical impedance measurements will be employed to select candidate standing wave operating frequencies for the water and clot experiments.

## Methods

### Transducer configuration

The transducer used for experiments was a radially polarized HCT with an outer diameter of 3.3 mm and an inner diameter of 2.5 mm, with a length of 2.5 mm, as employed in [28]. The transducer material was DL-47 (DeL Piezo Specialties, FL, USA). The transducer had a micro coaxial wire attached to the inner signal and outer ground electrodes using silver epoxy (Epotek H20E, Epoxy Technology Inc, MA, USA), and a catheter liner (PTFE Sub-Lite-Wall™ Liner 0.095" ID 0.0015" wall thickness, Zeus Company Inc, SC, USA) attached to its inner wall using epoxy (Epotek 301, Epoxy Technology Inc, MA, USA). This design is capable of achieving high internal pressures and generating cavitation clouds within the lumen of the transducer, as reported in previous work [28]. Finite element simulations of the transducer geometry were performed in OnScale™ as per the methods previously described in [28].

### Blood clot preparation

Clots were prepared similarly to previous sonothrombolysis studies [45–49]. Clots were prepared from pig blood extracted from the femoral vein of 70 kg pigs and captured in 3.2% sodium citrate vacutainers (BD Vacutainer™, Thermo Fisher Scientific Inc, MA, USA). Clots were formed in 2 mL borosilicate pipettes (Fisherbrand™ Disposable Borosilicate Glass Pasteur Pipets, Thermo Fisher Scientific Inc, MA, USA) after mixing anticoagulated blood with 100 mMol/L $CaCl_2$ at a ratio of 1 mL of anticoagulated blood to 200 µL of $CaCl_2$. The sealed pipettes were incubated in a water bath at 37 °C for 3 hours, and then transferred to a fridge at 4 °C for 3 days. The resulting retracted clots were extracted and cut into homogenous 5 mm lengths prior to experiments.

Cavitation observation with high-speed ultrasound imaging

   *a) Experimental configuration*

The experimental setup (Figure 1) consisted of a Vevo 2100 imaging system with a 40 MHz probe (MS550D) positioned to image the center plane of the transducer. The frame rate was set to 916 Hz, the system's maximum for a field of view that captured the entirety of the lumen. The imaging system was synchronised with treatment pulses using a pulse delay generator (Model 575, Berkeley Nucleonics Corp, CA, USA) to ensure every treatment pulse occurred at the same time point during each frame. The transducer was driven by a function generator (AFG3102, Tektronix, OR, USA) and an RF amplifier (A150, E&I, NY, USA). The voltage across the transducer was measured using a 100:1 voltage probe and digitized (PicoScope 5242B, Pico Technology, Cambridgeshire, UK) at a sample rate of 125 MS/s (14 bits). The applied voltage amplitudes reported are the voltages measured across transducer.

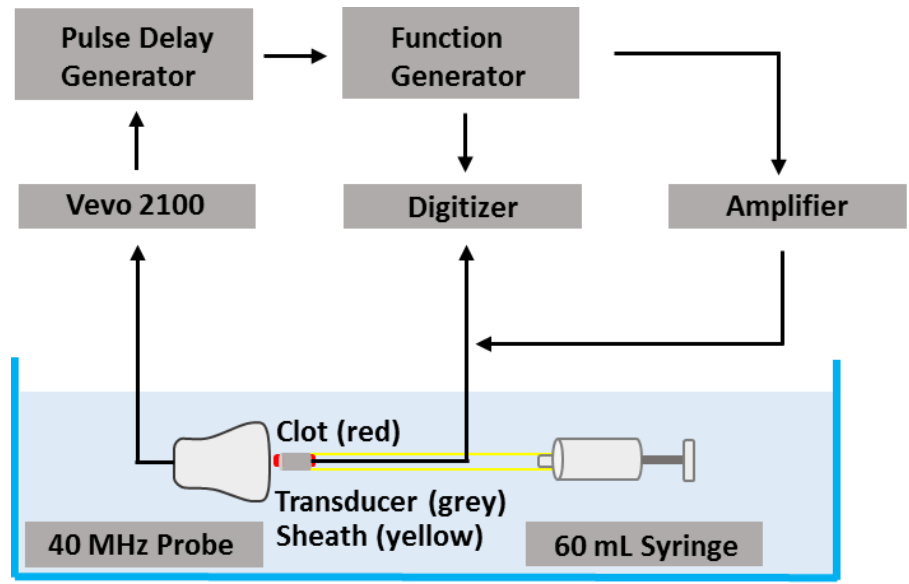

*Figure 1. Schematic overview of experimental configuration.*

   *b) Transmit frequency selection*

Experiments were performed in a tank of non-degassed deionized water held at 30 °C. The impedance of the transducer in water from 300 Hz to 8 MHz in 5 kHz steps was measured with a network analyser (AA-30.ZERO, RigExpert, Kyiv, Ukraine) with both water and a clot in the transducer lumen. Within the thickness mode bandwidth of the transducer there exist a number of impedance minima corresponding to standing wave frequencies of the geometry, which were previously shown to correspond to local pressure maxima [28]. To determine the optimal frequency for treatment, frequency sweeps (65 $V_p$ applied voltage, 10 µs pulses, 1000 Hz PRF) around each these local minima were performed under B-mode imaging. In [28] when higher voltage (82 $V_p$) was applied, cavitation was evident on imaging in the vicinity of three minima within the thickness mode bandwidth. Here, with a lower applied voltage, visible cavitation occurred at only one of the minima, which was the selected operating frequency for subsequent experiments. This was done with both water in the lumen and a clot in the lumen, resulting in different optimal frequencies for each case.

### c) Experimental protocol

For experiments with water in the lumen, the transducer was operated at 6.17 MHz, and pulse lengths of 10 µs and 100 µs were tested. 100 pulses were sent with a pulse interval of 10.92 ms such that one pulse was imaged every 10 frames for 1000 frames. Applied voltages from 19 to 82 $V_p$ were tested. This was repeated 3 times for a total of 300 treatment pulses at each voltage. We note that the maximum voltage applied here corresponds closely to that reported in [28] (82 $V_p$), which resulted in high levels of cavitation and did not cause any change in performance of the transducer over time.

For experiments with a clot in the lumen, a frequency of 6.43 MHz was used. A pulse length of 10 µs and a pulse interval of 1.09 ms was used for 916 pulses, resulting in a pulse every frame for 916 frames. A total of 5 clots were tested in this manner. For each clot, the treatment was performed 5 times at 19 $V_p$ and then once at 82 $V_p$. After treatment, the clots were extracted and bisected for optical imaging under a microscope.

### d) Cavitation threshold

B-mode imaging data was used to determine whether cavitation occurred on a given pulse. Each frame corresponding to a specific treatment pulse was manually tagged for the presence or absence of visible cavitation. To determine cavitation probability in water, for a given applied voltage, the ratio of number of pulses which had corresponding visible cavitation activity to the number of total pulses was calculated. This probability as a function of applied voltage was fitted to a sigmoid curve. The voltage value at a probability of 0.5 is defined as the cavitation threshold voltage.

## The self-sensing signal characterization analysis

For each pulse, recorded time traces were divided into a pre-trigger window, a window over the pulse transmit body, and a window of the ringdown post-transmit. The pre-trigger trace windows had a length equal to that of the main pulse body and were used as a baseline noise measurement in spectral analysis. The post-pulse window showed a ringdown of 1~3 µs after the transmit pulse body.

### a) Envelope

For signals in the time domain, the relative average amplitudes of the envelopes of the pulse body and the pulse ringdown can provide insight into signal features. The sum of the power of the envelope of an initial region of the transmit pulse (i.e. 1.2 µs in water and 1.67 µs with a clot) divided by that of the ringdown envelope (first 3 µs) provided a ratio of ringdown to body envelope amplitudes and served as a measure of cavitation quantification. Only this initial region of the transmit signal was used as this is before the acoustic wave front will reflect off the opposite inner surface wall [28].

### b) Spectra

For frequency domain analysis, Tukey windows with a cosine fraction of 0.25 were used for 100 µs pulses while Blackman windows with $\alpha = 0.16$ were used for 10 µs pulses. For 10 µs pulses, 10 times zero padding was used to make the window the same size as for 100 µs pulses. Fast Fourier transforms were performed for each set of pulses for both the pre-trigger and the pulse transmit windows and the spectra were averaged for each applied voltage. The pre-trigger spectra provided baseline noise measurements. As one means of quantifying cavitation, the sum of the power was calculated over the frequency windows surrounding the fundamental frequency (i.e. 6.028-6.339 MHz for 10 µs pulses and 6.15-6.2 MHz for 100 µs pulses in water), ultraharmonic frequency (i.e. 9.208-9.308 MHz for both pulse lengths in water and 9.554-9.674 MHz with a clot) and a frequency range corresponding to broadband signals (i.e. 19.3-19.7 MHz for 10 µs pulses in water; 5.13-5.4 and 7-7.3 MHz for 100 µs pulses in water; 5.451-5.598 and 7.45-7.7 MHz with a clot).

## Results

### a) Impedance and work frequency

The HCT impedance as a function of frequency in water is shown in Figure 2A. Local minima associated with the length and 3rd harmonic length resonant modes appear at 660 kHz and 1.93 MHz, along with a more complex series of peaks and troughs near the thickness mode bandwidth (4.9-7 MHz). Figure 2B shows measurements for both water and clot filled lumens. A number of local minima are seen around the thickness mode bandwidth spaced roughly 600 kHz apart. Each of these minima correspond to a standing wave frequency [28]. Having a clot in the lumen shifts the frequency minima's by ~260 kHz. The difference in transmit frequencies used for the water in the lumen and clot in the lumen cases (6.17 and 6.43 MHz) is consistent with this shift in the impedance curve.

Figure 3 shows the simulated pressure map and lateral pressure distribution for the transducer geometry operated at 6.17 MHz in water. These results show a narrow high-pressure region along the central axis of the cylinder and low-pressure regions near the transducer walls.

*b) Cavitation imaging and cavitation probability with water*

Figure 4 shows frames from B-mode imaging of the transducer operating in water. At low voltages, only water is visible in the lumen. As the voltage increases, cavitation clouds begin to appear, and at high voltages, there is strong cavitation activity all within the central lumen of the transducer. These qualities are similar for both 10 and 100 µs pulses, with 100 µs pulses showing a broader spatial distribution of the cavitation cloud at higher voltages.

Cavitation probability in water derived from B-mode imaging is shown in Figure 5. No cavitation was observed before 54 $V_p$ for both 10 and 100 µs pulses. For both pulse lengths, above 65 $V_p$ cavitation was observed on every pulse.

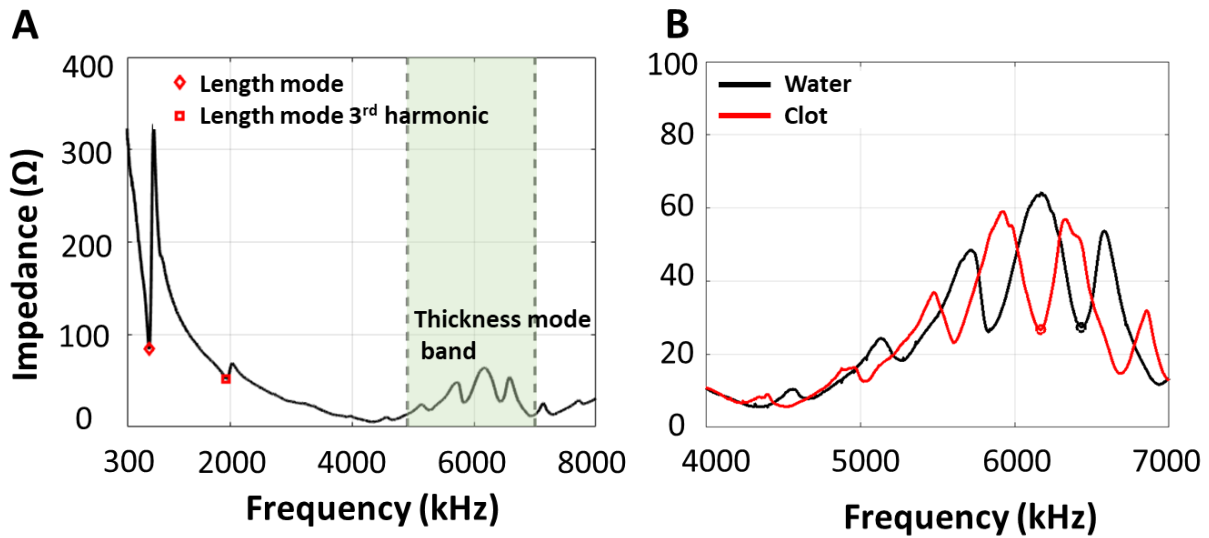

*Figure 2. The measured electrical impedances as a function of frequency in a range of 300 – 8000 kHz (A) and 4000 – 7000 kHz (B) with either water or clot occupying the transducer lumen. Minimas at 660 kHz and 1.93 MHz correspond to the length mode and its 3rd harmonic. The transmit frequencies used for clot treatments and treatments with water in the lumen are highlighted with circle markers on (B).*

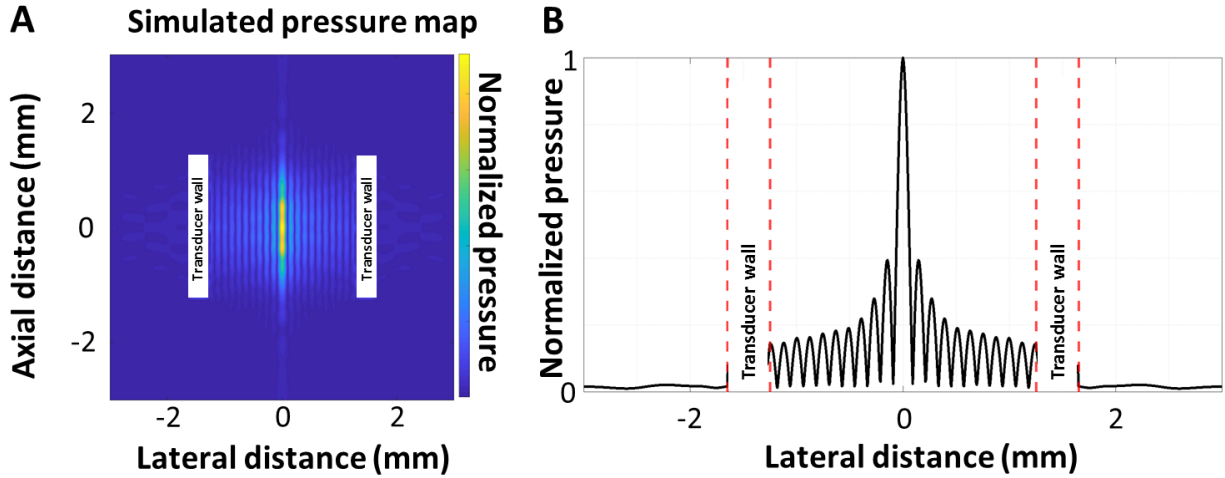

*Figure 3: Simulated pressure map (A) and pressure along dashed red line (B) for transducer operating at 6.17 MHz in water.*

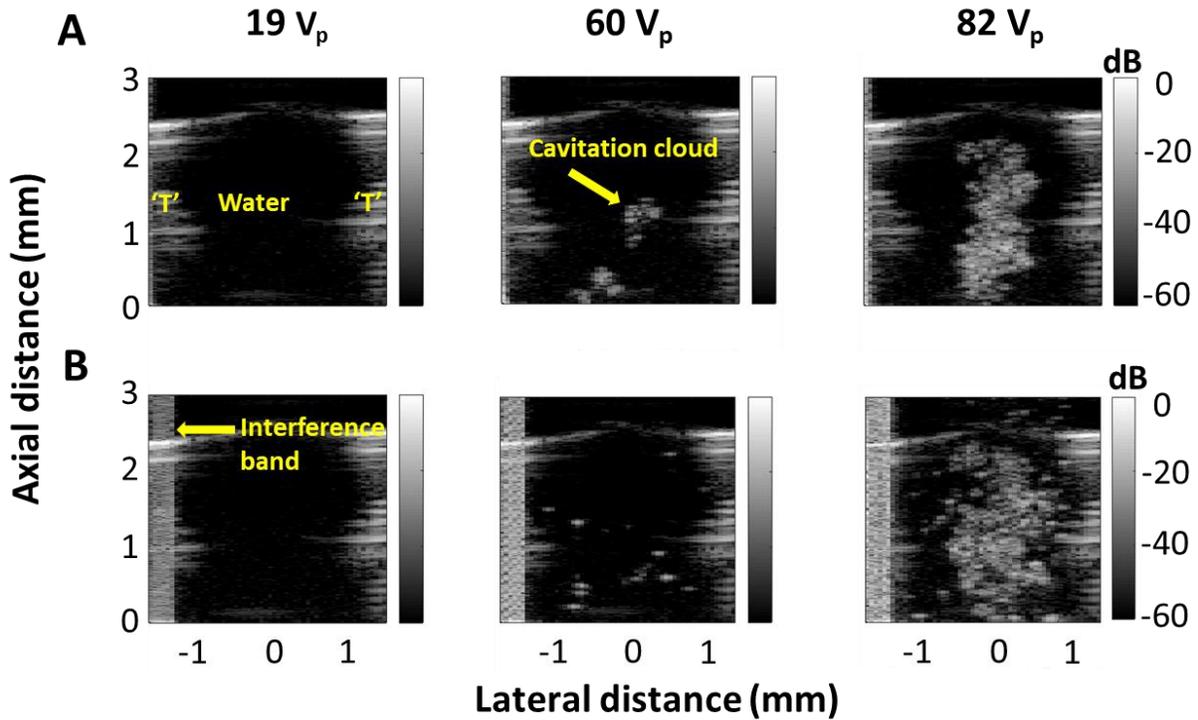

*Figure 4: Frames from B-mode imaging of lumen of transducer in water with 10 μs (A) and 100 μs pulses (B). 'T' represents transducer walls.*

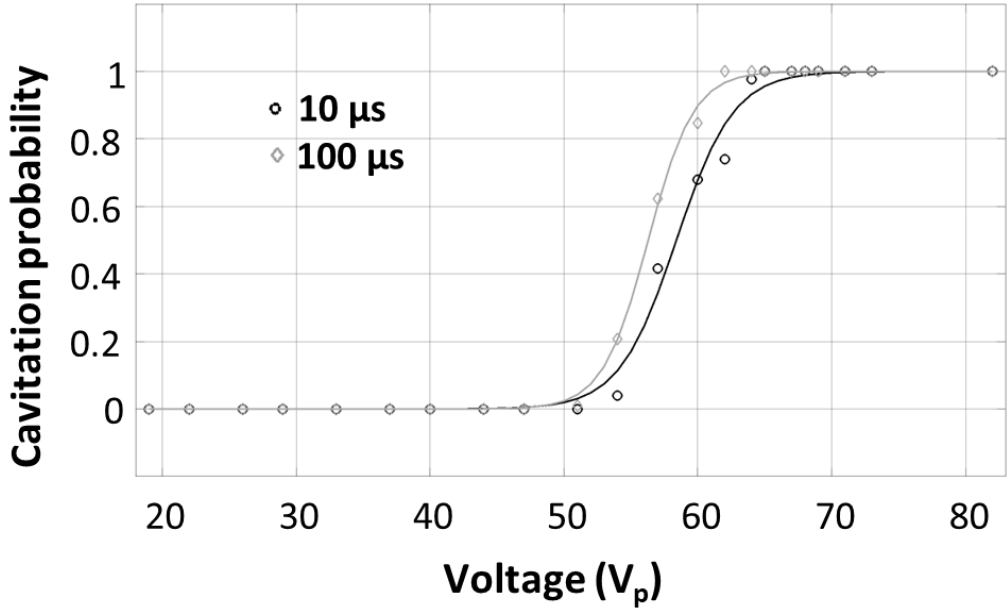

*Figure 5. Cavitation probability as a function of applied voltage and measured using B-mode imaging for both 10 and 100 μs pulses in water.*

### c) Self-sensing signals in water experiment

Figure 6 shows time-domain voltage signals for both 10 and 100 μs pulses and at 19, 60, and 82 $V_P$ in water. 19 V was shown to be significantly below the cavitation threshold as verified with B-mode imaging (Figure 3), while 60 $V_p$ is around the cavitation threshold voltage range, and 82 $V_p$ is significantly above the cavitation threshold where cavitation was visible on every pulse. Comparing the envelopes of the voltages at each applied voltage shows differences in the ratios of the amplitude of the envelopes of the ringdown portions to the pulse bodies. The relative ratios decrease from 19 to 60 to 82 $V_p$ as the ringdown portion of the envelope decreases in amplitude. This is consistent for both the 10 and 100 μs pulses.

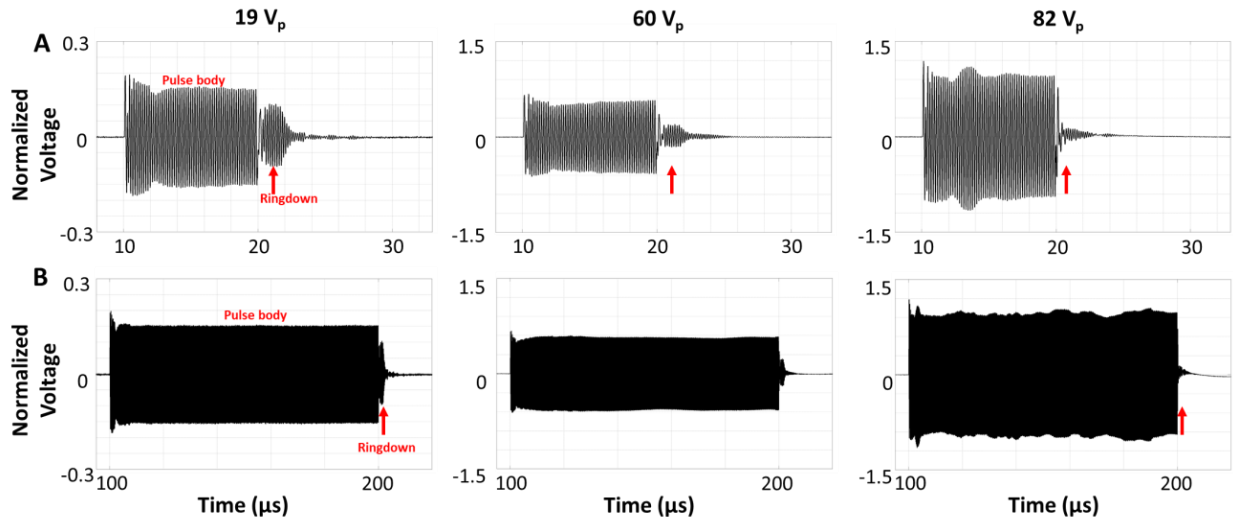

*Figure 6. Voltages measured across the transducer during (A) a 10 μs and (B) 100 μs pulses at different applied voltages in water. The red arrows point to the ringdown portions of the signals.*

Figure 7(A-B) shows the ratios of the amplitude of the ringdown portion of the signal envelope to the amplitude of the pulse body as a function of applied voltage in water. For both 10 and 100 μs pulses, at lower voltages below the cavitation threshold the ratio remains constant. As the cavitation threshold voltage is approached the ratio begins to decrease with increasing applied voltage. The standard deviation of measurements also begins to increase past this point. Figure 7(C-D) shows the normalized variance of these measurements as a function of voltage, showing an increase starting from the cavitation threshold voltage before plateauing at around 64 $V_p$ in the 10 μs case and 67 $V_p$ in the 100 μs case.

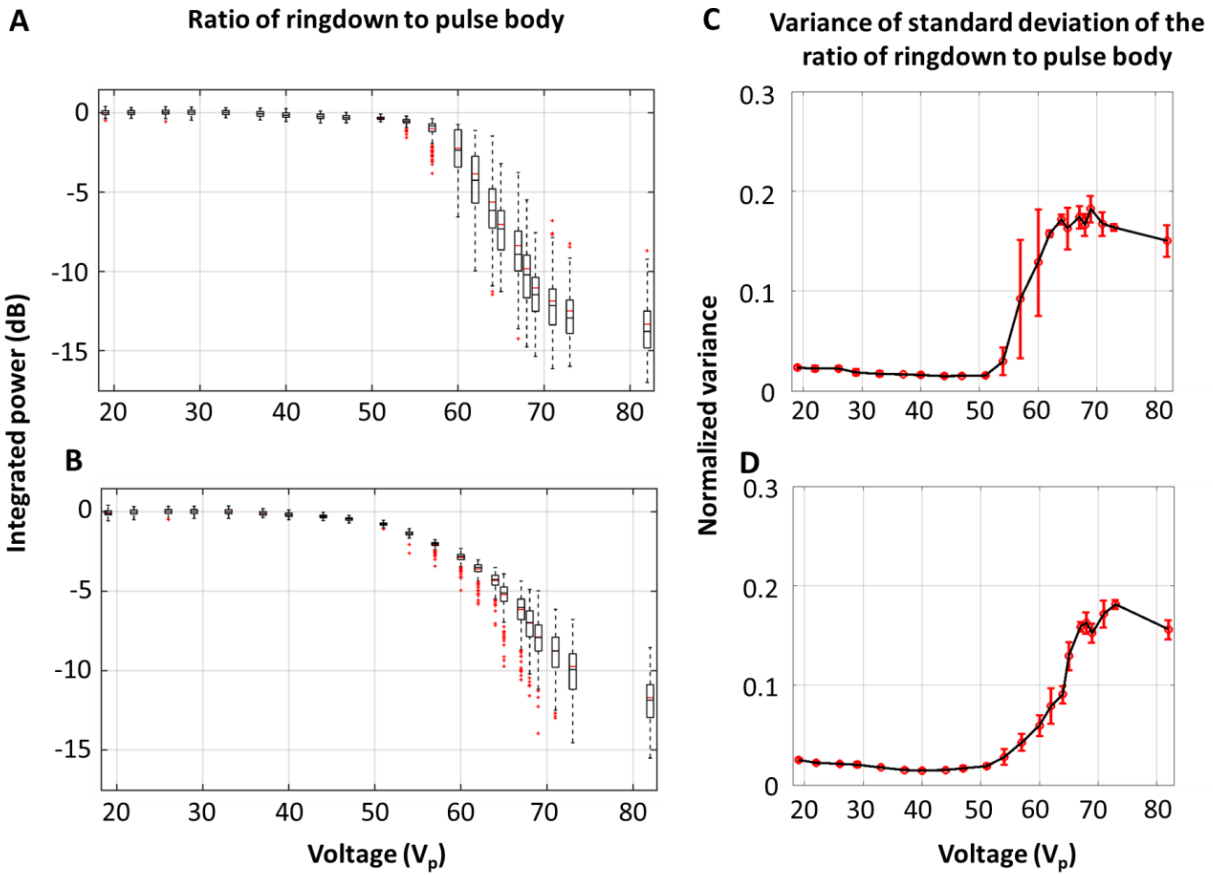

*Figure 7. Box plots of the ratio of the integrated power of the amplitude of the envelope over the ringdown portion of the signal to the body of the pulse as a function of voltage for (A) 10 μs and (B) 100 μs pulses in water, and the normalized variance of the same ratio as a function of voltage for (C) 10 μs and (D) 100 μs pulses. The error bars represent standard deviation.*

The recorded voltages across the transducer shown in the frequency domain reveal the emergence of harmonics as a function of applied voltage. The spectra of the pulse body for select voltages are shown in Figure 8, where ultraharmonics associated with inertial cavitation can be seen at voltage around and above the cavitation threshold voltage.

The quantification of spectra frequency bands as a function of applied voltage is shown in Figure 9. For 10 μs pulses, an increase in broadband noise is visible starting from 62 $V_p$, and continues to increase with applied voltage until approximately 69 $V_p$. For the ultraharmonic signal, a significant increase is notable from 47 $V_p$, and the power increases with voltage until beginning to level off around 54 $V_p$. The fundamental frequency signal increases monotonically as a function of applied voltage.

The general trends for the 100 μs pulses are similar, with broadband signals increasing from a slightly lower voltage of 57 $V_p$, ultraharmonics beginning to increase at 51 $V_p$, and the fundamental signal increasing with applied voltage with no threshold behavior.

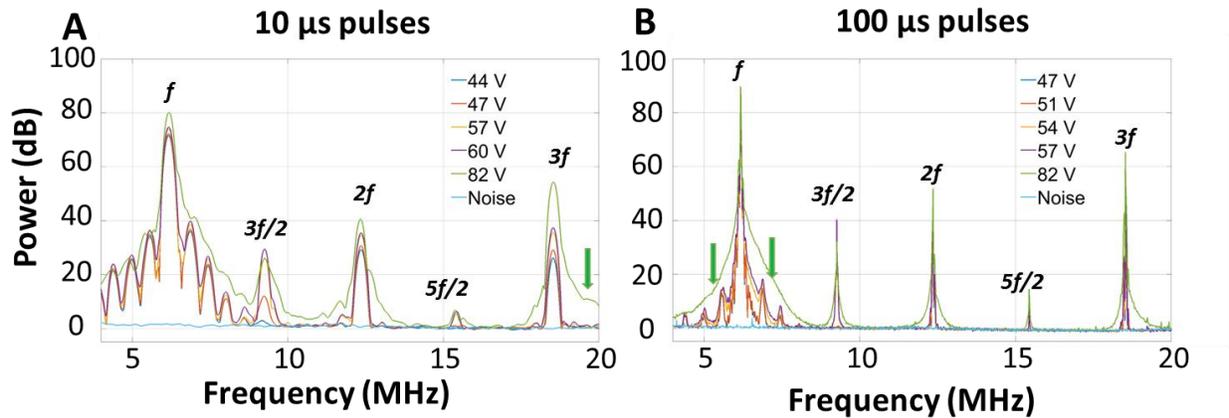

*Figure 8. The spectra of (A) 10 μs and (B) 100 μs pulses at different voltages in water. The green arrows point to regions of broadband emissions which are visible at 82 $V_p$.*

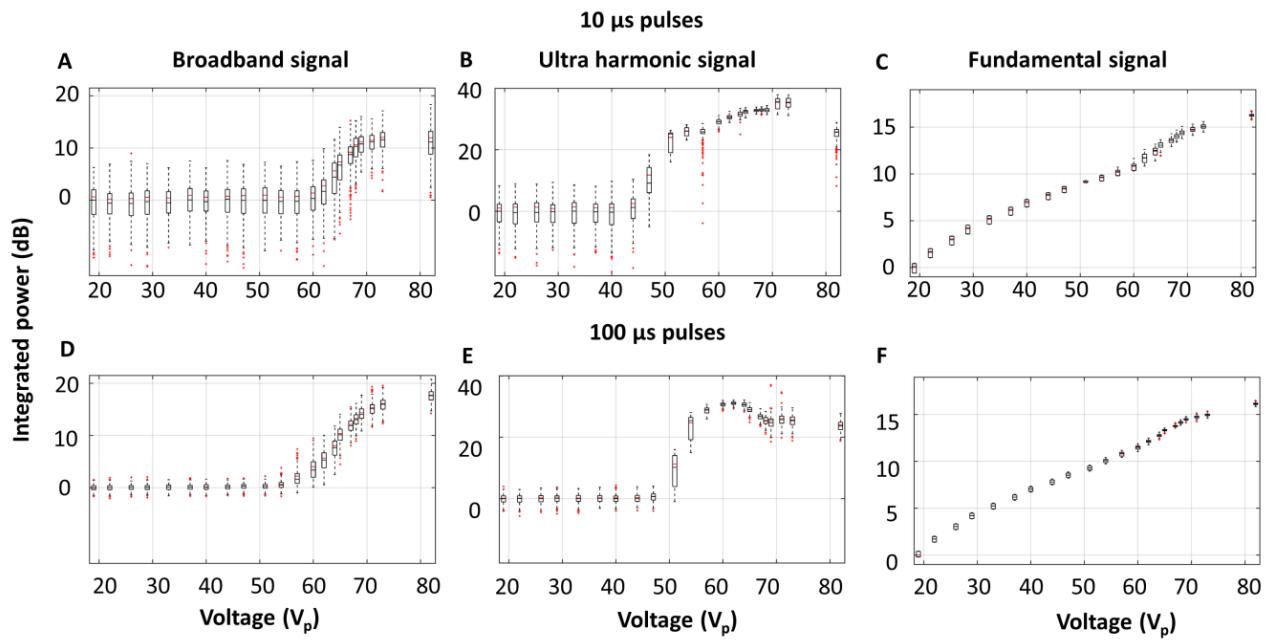

*Figure 9. Box plots of integrated power of specific frequency bands of the transducer voltage spectra as a function of applied voltage in water. The broadband (A, D) and ultraharmonic (B,E) for both 10 μs (A,B) and 100 μs (D,E) pulses show a threshold behavior, while the fundamental frequency range (10 μs pulses, C, and 100 μs pulses, F) show a monotonically increasing behavior.*

*d) Self-sensing signals in clot treatment*

Figure 10 shows representative B-mode images of a treatment at 82 $V_p$ with a clot in the lumen of the transducer. A central cavitation cloud is visible from the beginning of the treatment, and post-treatment a lesion in the same location is visible. The bisected extracted clot shows the lesion hole under optical imaging.

Figure 11A shows representative time-domain voltage signals during treatment of a clot captured at applied voltage of 19 and 82 $V_p$ along with their associated frequency-domain spectra. Similar to the signals of the transducer with water in the lumen, the relative amplitudes of the ringdown portion of the envelope to the pulse body portion decrease significantly from 19 to 82 $V_p$. The spectra shown in Figure 11B also show ultraharmonics present at 82 $V_p$ which are not visible at 19 $V_p$.

The quantification of these results across the 5 clots tested is shown in Figure 12 as violin plots. Using the envelope, broadband signals, or ultraharmonic signals all show significant (P<0.0001) differences in integrated power between 19 and 82 $V_p$.

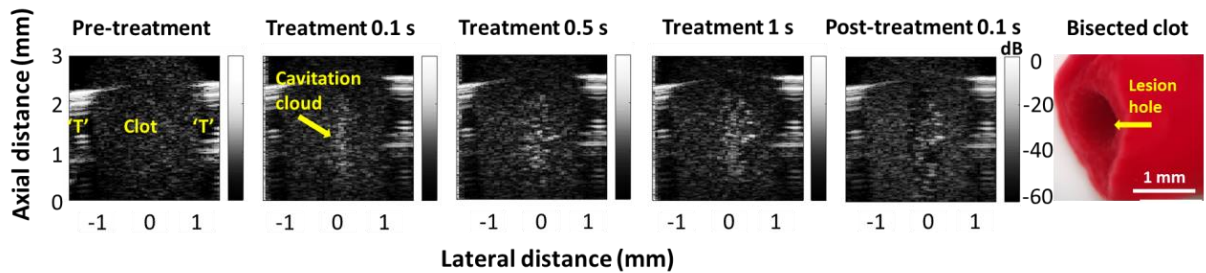

*Figure 10. Representative B-mode imaging of a cross section of a clot situated in the transducer before, during, and after treatment. 'T' represents transducer walls. Treatment parameters were 10 μs pulse length, 1000 Hz PRF, and 1 second treatment duration. Before treatment, the clot is visible as uniform speckle bounded by the side walls of the transducer. During treatment, the cavitation cloud is visible as a hyperechoic region along the central axis of the transducer. Immediately post treatment, the hypoechoic region visible over the same area of the cavitation cloud, indicating the generation of a lesion which is confirmed by the bisected clot.*

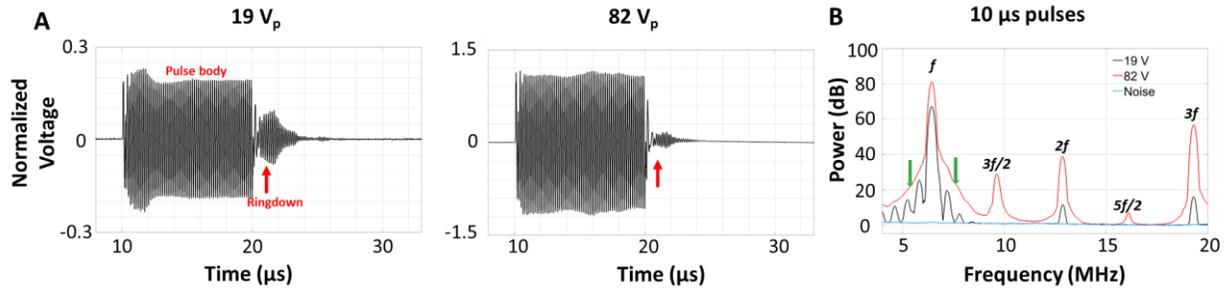

*Figure 11. Voltage signals across the transducer with a clot in the lumen at (A) 19 $V_p$ and (B) 82 $V_p$, and (C) the spectra of these signals. The red arrows on the time-domain traces indicate the ringdown portions of the signals, and the green arrows on the spectra indicate areas of broadband emissions at 82 $V_p$.*

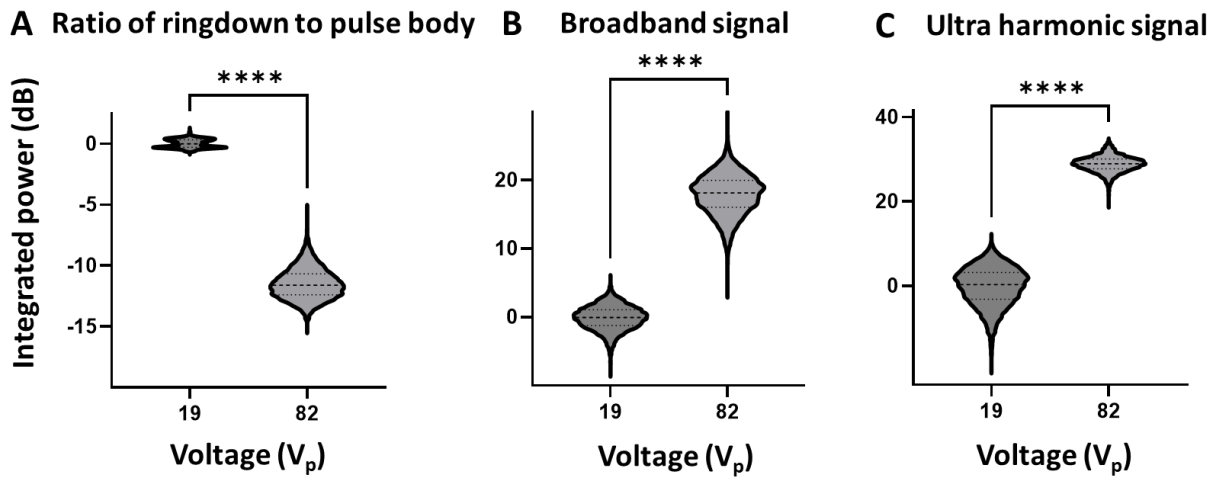

*Figure 12. Violin plots comparing the integrated power for (A) the ratio of ringdown to pulse body of the time domain envelopes and (B,C) of specific frequency bands of the transducer voltage signal spectra at both 19 and 82 $V_p$ with a clot in the lumen. \*\*\*\* indicates statistical significance between groups with $P<0.0001$ using two-tail, unpaired T-test.*

## Discussion

The results have collectively demonstrated the use of measured voltage signals across an HCT to detect cavitation induced within its lumen. The measured signals underwent both temporal and spectral analyses which were conducted by partitioning the signals into pulse body and ring-down components. For water-filled lumens, fluctuations in the envelope of the pulse body became evident following the onset of cavitation, which is consistent with previous reports for spherically focused transducers in the setting of boiling histotripsy conditions [38,50]. These fluctuations arise from the superposition of signals associated with cavitation (emissions, reflections/scattering) and the applied transmit signal voltage. It was also observed that the ratio of the ringdown to body amplitude remained relatively flat at lower voltages (~< 50 $V_p$), but after the onset of detected cavitation, it began to drop precipitously. In [28], the simulated and hydrophone-measured pressure waveforms external to the transducer exhibited a stepped decay over time following the cessation of the stimulating pulse, which was attributed to the dissipation of internally reflected cylindrical waves. The measured ringdown signals are likely predominantly associated with internally reflecting waves that dissipate over time. The reduction in amplitude of the ringdown ratio with increasing levels of cavitation clouds is hypothesized to result from interaction of the internally reflected waves with the clouds. That is, it can be expected that waves encountering clouds will be scattered and reflected, and lose energy through absorption processes. The reduction in self-sensing detected amplitudes – which arise from waves impinging on the luminal surface – may also be further reduced by cloud induced factors such as propagation velocity changes that alter interference patterns. In terms of the pulse body spectral analysis, at lower voltages below the imaging detected cavitation threshold, the spectra feature a prominent peak at the transmit frequency alongside a number of smaller adjacent peaks within the thickness mode bandwidth. Multiple harmonics (2nd, 3rd) are also present, consistent with the occurrence of nonlinear propagation, possibly with additional contributions from nonlinear transducer behavior and electronic behavior. The onset of detected ultraharmonic signals (3/2 and 5/2), a pressure threshold-dependant bubble behavior, is consistent with the presence of cavitation and corresponds to the onset voltage range for cavitation detected with imaging. The detection of broadband signals followed a similar pattern, albeit with a higher voltage onset and variance. In interpreting the 'broadband' signal, it should be noted that the frequency ranges selected for quantification were within the thickness mode range (outside of local peaks) and adjacent to the 3rd harmonic signal peak. In [28], hydrophone measurements performed adjacent to the lumen at higher voltages demonstrated broadband cavitation across a wide range of frequencies. This was not observed in the self-sensing HCT spectra, which is attributed to the frequency response of the transducer when exposed to emissions impinging on its lumen surface. As the frequency ranges employed were in proximity to the fundamental and 3rd harmonic peaks, there may be contributions from spectral broadening of these peaks in addition to general broadband cavitation that can be associated with inertial cavitation.

These experiments with water filled lumens were carried out at an operating frequency selected based on maximizing cavitation. As observed in [28], the frequency corresponded to one of the impedance minima frequencies, which are associated with standing wave frequencies where the effective gain is highest. The location of impedance minima was found to differ between water and clot filled lumens (Figure 2B) due to differences in the compressional wave propagation velocity: clots have a higher wave speed than water [51], which in turn shifts the frequency of the minima. In a clinical setting, clots can have a range of compositions and ages, which can be expected to give rise to variations in the propagation velocity. This suggests than impedance measurements will be important to incorporate into the treatment process. The proof of principle clot experiments was conducted at two voltages - one below the cavitation threshold and one at the maximum voltage. The high voltage exposures resulted in the production of a lesion within the clots, as observed in our recent study [28]. The cavitation and lesions were preferentially initiated in the central portion of the lumen, consistent with the higher on-axis pressure profile (Figure 3). No cavitation was present lateral to the transducer, due to the low-pressure levels. The average spectra during the course of the treatment had similar features to the water-filled lumens. A comparison of the ringdown, ultraharmonic and broadband signals between the two voltages shows highly significant differences. This supports that these are candidate metrics to assess for the presence of cavitation during clot treatments.

The results of this work support that both electrical impedance and self-sensing signals of an intravascular scale HCT can provide information that is potentially relevant to the monitoring and control of an ultrasound enhanced aspiration. Shifts in impedance minima frequencies may provide a means by which to detect when a clot is ingested into the transducer lumen, which is information that is relevant to determining when to commence the treatment scheme. The impedance minima can also be used to identify candidate operating frequencies, both at the outset and during treatments. Cavitation detection metrics (e.g. ringdown ratio or selected spectral bandwidths) can further be used in selecting a particular operating frequency. An examination of the evolution of these signals over time during treatments – both for 'static' and aspirating clots – is warranted to determine their possible role in dynamically adapting exposures during the course of treatments. The simplicity and cost-effectiveness of the self-sensing and impedance approaches carry notable advantages, as they do not require additional detector transducers.

## Acknowledgement

This work was funded by NSERC Discovery and CIHR Project grants. The authors thank Jennifer Barry, Shawna Rideout, Dallan McMahon and Stephanie Furdas for performing animal blood collection.